\documentclass[Journal]{IEEEtran}
\usepackage{amssymb}
\usepackage[cmex10]{amsmath}

\usepackage{amsthm}
\usepackage{amssymb}
\usepackage{graphicx,tikz} 
\usetikzlibrary{arrows,automata}
\usepackage{balance}

\usepackage{array,cite} 
\usepackage{subfigure,epsfig} 

\title{Capacity of Diffusion-based Molecular Communication with Ligand Receptors}

\author{ Arash Einolghozati, Mohsen Sardari, Faramarz Fekri\\
School of Electrical and
Computer Engineering, Georgia Institute of Technology, Atlanta, GA 30332\\
\texttt{Email:}\{einolghozati, mohsen.sardari, fekri\}@ece.gatech.edu\vspace{-0.2in}
\thanks{This material is based upon work supported by the National Science Foundation under Grant No. CNS-111094}
}

\begin{document}

\maketitle
\thispagestyle{empty}
\pagestyle{empty}

\begin{abstract}
A diffusion-based molecular communication system has two major components: the diffusion in the medium, and the ligand-reception. Information bits, encoded in the time variations of the concentration of molecules, are conveyed to the receiver front through the molecular diffusion in the medium. The receiver, in turn, measures the concentration of the molecules in its vicinity in order to retrieve the information. This is done via ligand-reception process. In this paper, we develop models to study the constraints imposed by the concentration sensing at the receiver side and derive the maximum rate by which a ligand-receiver can receive information. Therefore, the overall capacity of the diffusion channel with the ligand receptors can be obtained by combining the results
presented in this paper with our previous work on the achievable information rate of molecular communication over the diffusion channel.
\end{abstract}
\section{Introduction}
\label{sec:intro}
Molecular diffusion plays a key role in different forms of communication in nature. At the microorganism scale, molecular signals are used for communication and control among cells. For example, one of the well-known communication primitives among cells is a phenomenon called \emph{Quorum Sensing}. Quorum sensing is a communication process among cells in which bacteria monitor the population density of their type (and possibly other types of bacteria) via the production and detection of special molecules. Quorum sensing allows bacteria to synchronize the behavior of the group, and thus act as a unit~\cite{Bassler1999,Bassler2002,Mehta2009}.

The communication can also appear in other forms such as \emph{pheromone} communication and \emph{calcium signaling}. Pheromone communication is performed using specific type of molecules released by plants, insects and other animals which trigger specific behaviors among the receptor members of the same species. Pheromones of a particular type, after being released into the environment, diffuse in the air until they are captured by the same receptor type.
Cells absorb calcium molecules in response to various stimuli that open/close particular channels on the cell membrane. The molecular information in the variation of the calcium ions concentration is propagated both inside and outside the cell, causing a variation in the electrical charge of the cell membrane and, subsequently, the transduction of the information into an electrical signal.

Due to various limitations such as size, arguably the dominant form of communication in nano scales is via molecular signaling, which is fundamentally different from conventional electromagnetic-based communication.
Recent advances in bio-nano technology has motivated research on designing nanoscale devices to perform tasks similar to their biological counterparts. This opens up several potential applications. For example, it could lead to molecular based networks that can be deployed over or inside the human body to monitor glucose, sodium, and cholesterol, to detect the presence of different infectious agents, or to identify specific types of cancer. Such networks would also enable new smart drug administrative systems to release specific drugs inside the body with great accuracy and in a timely manner. To enable all such applications, the communication is the key.

The wide range of potential applications has inspired development of new theoretical frameworks for molecular communication~\cite{eckford2010,Eckford2008,Kadloor2009,Pierobon2010}. The focus of these studies is to encode the information into the time of the release of molecules while the propagation from the transmitter to the receiver is governed by Brownian motion (without drift~\cite{Eckford2008} and with drift~\cite{eckford2010,Kadloor2009}). Although Brownian motion of particles is basically the same process as diffusion, natural organisms do not look directly at the Brownian motion of a large particle. Instead, they measure diffusion of small molecules by following changes in concentrations.
Therefore, we consider an alternative molecular communication system governed by diffusion process in the medium, referred as the \emph{diffusion channel}~\cite{ISIT2011_Arash}. Arguably, the most dominating form of the communication at the micrometer scale is diffusion based molecular communication, i.e., embedding the information in the alteration of the concentration of the molecules and rely on diffusion to transfer the information to the destination. The communication among bacteria, calcium signaling, and many others can be reduced to diffusion based molecular communication. Another body of work in this field involves the study of the Quorum Sensing as a network and mapping the Quorum Sensing to consensus problem under diffusion-based molecular communication~\cite{CISS2011_Arash}. 

In our previous work~\cite{ISIT2011_Arash}, we studied the simpler and more practical discrete systems. There, we studied the achievable rate over the diffusion channel when we are limited to the amplitude modulation (on-off keying), i.e., a discrete diffusion channel. One of the important problems that where left open in~\cite{ISIT2011_Arash} was the reception process. As explained later, the dynamics of the reception in molecular communication is fundamentally different from the conventional receivers. As a result, the receiver can limit the capacity of the molecular communication. In this work, we investigate the dynamics of the molecular receptor and its effect on the capacity of the molecular diffusion communication. Further, using the results in this paper, we verify some of the basic assumptions we made in our previous work.

We consider a communication scenario consisted of a transmitter and a receiver communicating via the diffusion channel, as shown in Fig.~\ref{fig:model}. In this context, the transmitter is capable of releasing molecules to the medium to change their concentration according to information bits~\cite{ISIT2011_Arash}. Similarly, the receiver is capable of absorbing the molecules or chemical signals, for example, by using ligand-receptor binding which is a trans-membrane receptor protein on a receiving cell~\cite{Keramidas2004,Model1995}. Ligand-receptors provide a bin for molecules to bind to. This process triggers a signaling inside the cell indicating the reception of the molecule. The receiver has a large number of binding places using which it can estimate the concentration by averaging over all binding places. The performance of the ligand-receptor has a profound impact on the reception of information. Once a molecule is bound to a receptor, it takes some time for the receptor to reset, we later address this in more detail. Our goal is to find the \emph{capacity} in such a diffusion-based molecular communication channel using the models we develop for the ligand-receptor. 

 The rest of the paper is organized as follows: In Sec.~\ref{sec:background}, we review the model for diffusion-based molecular communication. In Sec.~\ref{sec:model}, the model for an ideal receiver and its capacity is studied and then it is extended to a Markov-based model for receptors. Sec.~\ref{sec:conclusion} addresses the link between this paper and our previous work and concludes the paper.

\section{Molecular Communication Model}
\label{sec:background}

\begin{figure}[t]
\vspace{-.35 in}
\hspace{-1in}
\centering
\includegraphics[height=0.8\linewidth, angle=-90]{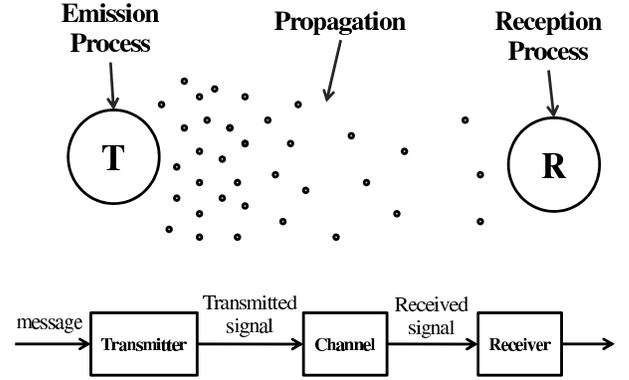}
\vspace{0.4in}
\caption{ Basic molecular communication model: the transmitter,diffusion medium, and the receiver.}
\label{fig:model}
\end{figure}

The paradigm for diffusion-based molecular communication is depicted in Fig.~\ref{fig:model}. We assume a source with binary alphabets. The information bits are encoded into alteration of the rate of molecule production by the transmitter {\bf T}. In particular, we consider pulse-shaped modulation, i.e., the transmitter either releases molecules over a time period or remains silent. The emitted molecules are transferred through the medium by diffusion process and reach the receiver {\bf R}. The flow of molecules from the regions with high concentration (around {\bf T}) to the regions with lower concentration (around {\bf R}) is governed by the diffusion equation. The second Fick's law of diffusion, as discussed in~\cite{ISIT2011_Arash}, gives the concentration of molecules at each position in the medium over time, proportion to the input rate. We can view the diffusion in the channel as a low-pass linear filter which takes the production rate as the input and gives the concentration at the receiver as the output via convolution with the impulse response~\cite{ISIT2011_Arash}.
The low-pass frequency response of the diffusion channel restricts the rate of the information exchange; the receiver does not observe the high frequency elements which in turn means that the rapid changes in concentration of molecules at te receiver are absent. In~\cite{ISIT2011_Arash}, we also observed that there is an inherent memory in the medium. Molecules tend to linger in the medium after each transmission and some amount of time is needed for the molecules to diffuse away and channel to be reset. In~\cite{ISIT2011_Arash}, we considered all the above constraints and obtained the capacity of the diffusion channel; assuming a pulse amplitude modulation (of the molecule production).

\begin{figure}[t]
\centering
\hspace{-.8in}
\includegraphics[height =0.8\linewidth, angle = -90]{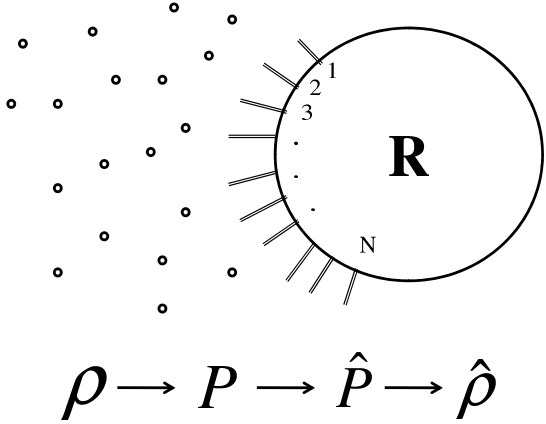}
\vspace{0.5in}
\caption{Illustration of the ligand receptor mechanism for sensing the concentration.}
\label{fig:Receiver}
\end{figure}

Once the diffusion process transfers the molecules to the receiver, the receiver {\bf R} measures the concentration of molecules, using a set of \emph{binding sites} that become active once they trap a molecule. Each binding site can trap one molecule at a time and cannot accept any more until the reaction to the previously trapped molecule is completed. The model for the receptor is shown in Fig.~\ref{fig:Receiver}. As the concentration $\rho$ increases, the probability $p$ that a molecules is trapped and hence the number of active sites increases. This way, the receiver can estimate the concentration in the medium. Note that the receiver samples these binding sites (for being in active state) synchronously at the end of each bit transmission (as pulse-shape keying PSK in the conventional communication systems). The number of active binding sites will determine the received concentration. The information can then be extracted from the variation of the concentration of molecules at {\bf R}.


%
 
As mentioned earlier, the entire diffusion-based molecular communication system has two major components: the diffusion in the medium, and the ligand-reception. The capacity due to the diffusion channel is obtained in~\cite{ISIT2011_Arash}. In this paper, we study the fundamental limits on the information rate due to the ligand-reception. The overall capacity of the system, which is the maximum conveyed information per channel use, can be easily obtained by having the capacities of the individual components.

\section{Modeling the Ligand Receptors}
\label{sec:model}
Assume a receiver equipped with $N$ binding sites/receptors. In every moment, a receptor is either active (a molecule is trapped in it) or inactive (empty). We assume all the receptors are identical, observe the same concentration, and act independently. The process of determining the concentration of molecules at the receiver is as follows: the molecules arriving from the channel make up a time varying concentration $\rho(t)$ at the receiver vicinity. The higher the concentration around the receiver, the higher the probability that a molecule gets trapped in a receptor. We denote by $p$ the probability that a single molecule being bound to an empty ligand-receptor when the concentration $\rho(t)=\rho$. In other words,
\begin{equation}
\label{eq:probability}
p=\mathbf{P} \left(\text{Binding}\; | \; \rho   ~\&  \text{ the receptor is empty} \right)
\end{equation}
The function that maps the concentration of molecules to the probability of the absorption is assumed to be a monotonically increasing one, $[0,\infty] \rightarrow [0, 1]$. As discussed later, $p\in[0,1]$ is a realization of a random variable $P$.

To each receptor, we associate an indicator random variable $X_i,\; i \in \left\{1,2,...,N\right\}$, where $X_i$\rq{}s are Bernoulli distributed with parameter $p$, i.e., $\mathbf{P}(X_i=1)=p$. Let $X^{(N)}$ be the vector containing the outputs of all the receptors. The vector $X^{(N)}$ is the input used to estimate $p$. It is known that $\sum_{i=1}^N X_i$ is the sufficient statistics for estimation of $p$. This implies that the receiver functionality is to add up the values of all receptors at every sampling instant. Let $\hat{P}$ be the best unbiased estimator for $p$. Since $\hat{P}=\frac{1}{N}\sum_{i=1}^N X_i$, the expected value and variance of $\hat{P}$, given $p$, can be easily derived as $\mathbf{E}[\hat{P}]=p$ and $\text{Var}[\hat{P}]=\frac{1}{N}p(1-p)$.

Clearly, $\hat{P}$ depends on the value that ${P}$ takes. In other words, we have
\begin{equation}
\label{eq:conditional}
\mathbf{P}\left(\left. \hat{P}=\frac{i}{N}\right|P=p\right)={N\choose i} p^i (1-p)^ {N-i}.
\end{equation}
We use the shorthand $\mathbf{P}(i|p)$ for $\mathbf{P}\left(\hat{P}=\frac{i}{N}|P=p\right)$ in~\eqref{eq:conditional}. Hence, by computing $\hat{P}$, the receiver has an estimate for $p$ which in turn is a function of $\rho$. The receiver maps back the estimate $\hat{P}$ to a level of concentration $\hat{\rho}$. Thus, we have a chain of processes as shown in Fig.~\ref{fig:Receiver}. 
We observe that except for the relation between $P$ and $\hat{P}$, all the other relations are deterministic and one-to-one. Therefore, to study the capacity of the ligand receptor, it is sufficient to consider only the mutual information between $P$ and $\hat{P}$. This implies that
\begin{equation}
\label{eq:mutual}
C =\max_{f_{P}(p)} I(\hat{P};P)= \max_{f_{P}(p)} [H(\hat{P})-H(\hat{P}|P)]
\end{equation}
where $C$ is the capacity due to the ligand reception and $f_{P}(p)$ is the distribution function of $P$. In addition, $H(\hat{P})$ is the entropy of discrete random variable $\hat{P}$ which takes values from the set $\left\{\frac{i}{N},\; i \in{0,1,...,N}\right\}$. Using~\eqref{eq:conditional} and the definition of mutual information we get
\begin{equation}
\label{eq:mutual_capacity}
C= \max_{f_{P}(p)} \sum_i \int_p \mathbf{P}(i|p) \log \left( \frac{\mathbf{P}(i|p)}{\int_p \mathbf{P}(i|p)  f_P(p) dp} \right) f_P(p) \; \text{d}p
\end{equation}
Therefore, the problem is to obtain $f_{P}(p)$ that achieves the capacity in~\eqref{eq:mutual_capacity}.

The capacity of the receiver depends on the model assumed for functionality of the receptors. In the following, we study two models to describe the process of estimating the concentration of molecules. First, the ideal receptor model is discussed. Then, we extend this model to capture the more complicated functionality of the receptors using a Markov chain model.

\subsection{Ideal Receptors}
\label{sec:ideal}
\begin{figure}
\includegraphics[width=.95\linewidth]{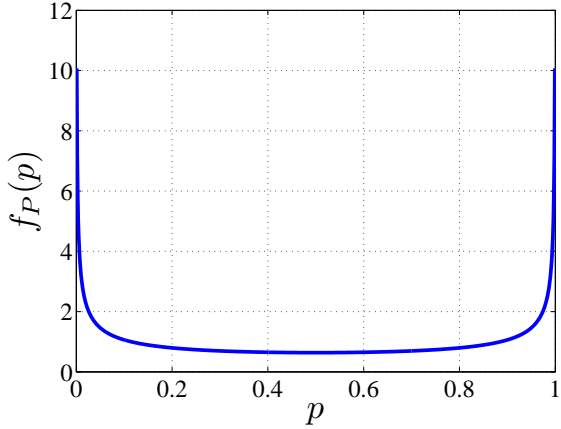}
\caption{Capacity-achieving distribution for $p$}
\label{fig:density_simple}
\end{figure}

\begin{figure}
\includegraphics[width=.95\linewidth]{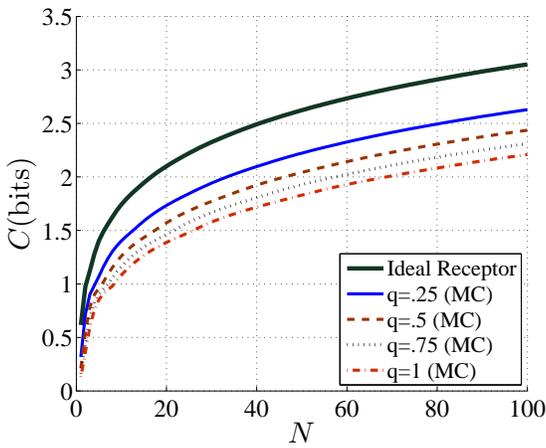}
\caption{Capacity versus the number of receptors}
\label{fig:capacity_plot}
\end{figure}

An ideal receptor is defined as a receptor which is always ready to receive molecules. In this model, molecules are assumed to occupy the receptors for a negligible amount of time. Hence, we assume that the molecules leave the receptor instantly after absorption and the condition on $p$ that needs the receptors to be empty is always satisfied. This means that the parameter $\hat{P}$ would be directly an estimate for the probability in~(\ref{eq:probability}). Therefore,  
\begin{equation}
\label{sufficient}
I\left(P;\hat{P} \right)=I\left(P;\textstyle  \sum_{i=1}^N X_i\right)=I\left(P;X^{(N)}\right).
\end{equation}
In this case, the problem of finding the capacity achieving distribution in~(\ref{eq:mutual}) resembles the corresponding problem in source coding which is discussed in~\cite{Minimax}. There, the problem is to find the distribution that maximizes the mutual information between an observed sequence and the unknown source parameter $\theta$. It is shown that the capacity achieving distribution is Jeffreys Prior which is proportional to the square root of the determinant of the Fisher information $\mathcal{I}$:
\begin{equation*}
f_P(\theta)\propto \sqrt{\mathcal{I}(\theta)}
.\end{equation*}
In our setup, we intend to maximize the mutual information between the sequence $X^{(N)}$ and the unknown parameter $p$. By comparison to the source coding problem we have:
\newtheorem{lem1}{Lemma}
\begin{lem1}
The capacity achieving distribution on $P$ that maximizes the mutual information $I(P,X^{(N)})$ follows the Jeffery's prior~\cite{Minimax}.
\end{lem1}
It is easy to verify that the Fisher information contained in $N$ independent Bernoulli trials with parameter $\theta$ is 
$$\mathcal{I}(\theta) = \frac{N}{\theta(1-\theta)}.$$ 
Therefore, the Jeffrey's prior for a Bernoulli random variable is given by the Arcsine distribution:
\begin{equation}
\label{eq:arc}
f_\Theta(\theta)=\frac{1}{\pi\sqrt{\theta(1-\theta)}}, \quad 0<\theta<1.
\end{equation}
This capacity achieving distribution is shown in Fig.~\ref{fig:density_simple}. The resulting distribution for $P$ shows that, in order to achieve the capacity, the transmitter should produce molecules in a way that approximately, makes $P$ close to either $1$ or $0$, in each transmission. This would also justify the use of a binary level pulse transmitter in our previous work~\cite{ISIT2011_Arash}. Intuitively, when $p$ is chosen to become closer to $\frac{1}{2}$, the variance $p(1-p)$ of the Bernoulli output of each receptor is increased. This would reduce the mutual information in~\eqref{eq:mutual}.
The numerical results for the capacity of the ligand receptors versus the number of receptors $N$ is shown in the outermost plot in Fig.~\ref{fig:capacity_plot}. Expectedly, the capacity is monotonically increasing with respect to $N$. 

\vspace{-0.08in}
\subsection {Markov Chain Model for Receptors}
\label{sec:markov}
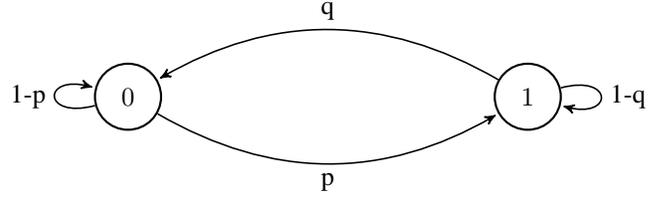
\begin{figure}
\centering
\begin{tikzpicture}[->,>=stealth',shorten >=1pt,auto,node distance=.6\linewidth,
semithick]
\tikzstyle{every state}=[fill=white,draw=black,thick,text=black,scale=1]
\node[state]         (A)              {$0$};
\node[state]         (B) [right of=A] {$1$};
\path (B) edge  [bend right] node[above] {q} (A);
\path (A) edge  [bend right] node[below] {p} (B);
\path (B) edge  [loop right] node {1-q} (B);
\path (A) edge  [loop left] node {1-p} (A);
\end{tikzpicture}
\caption{Markov chain model of  a ligand receptor}
\label{fig:markov}
\vspace{-0.1in}
\end{figure}

In~\cite{Relay_biochemical}, a more realistic model to address the lingering of molecules in receptors is used. In order to improve our ideal receptor model, we consider a Markov Chain (MC) model to better capture performance of receptors. This is due to the fact that a molecule which is trapped by a receptor does not leave instantly and will occupy the binding site for a random amount of time. During this interval, the receptor remains occupied and no other molecule is able to bind to the receptor. Hence, the probability that each receptor accepts a molecule is no longer $p$. Instead, some randomness would exist because of the non-ideal performance of the receptors. The MC model for each receptor is shown in Fig.~\ref{fig:markov}. In this figure, state $0$ corresponds to an empty receptor, i.e., $X_i=0$, whereas state $1$ corresponds to the receptor with a trapped molecule, i.e., $X_i=1$. Further, $p$ is the same as the previous section defined in~\ref{eq:probability}, and $q$ is the probability that the trapped molecule is released at each step of the MC.
%
The steady state probability distributions for this simple MC are given by 
\begin{equation*}
\left\{
\begin{array}{lcl}
\pi_0	&=&	\mathbf{P}(X=0)=\frac{q}{p+q}\\
\pi_1	&=&	\mathbf{P}(X=1)=\frac{p}{p+q}
\end{array}
\right.
.\end{equation*}
Since the diffusion channel is low-pass, the variations in the channel are slow relative to the mixing time of the MC, which is the time that takes for the MC to reach steady state. In other words, the transition time to reach the steady state is quite small compared to the time intervals that information is transmitted via the diffusion channel. Therefore, we may assume that the MC reaches the steady state before the concentration level at the receiver switches from high to low or vice versa. 

Let $\hat{\Pi}_1$ be the estimator for $\pi_1$ and denote by $\Pi_1$ the random variable representing $\pi_1$. Again, we have:  $\hat{\Pi}_1=\frac{1}{N}\sum_{i=1}^N X_i$. We use the method in the previous section to obtain the optimal distribution for $\Pi_1$ to maximize the mutual information described in below:
\begin{equation}
\label{eq:mutual2}
C =\max_{f_{\Pi_1}(\pi_1)} I(\hat{\Pi}_1;\Pi_1)= \max_{f_{\Pi_1}(\pi_1)} [H(\hat{\Pi}_1)-H(\hat{\Pi}_1|\Pi_1)]
\end{equation}
However, the difference is that $\pi_1$ is limited to $[0 ,\frac{1}{1+q}]$ as $p$ varies in the range $[0,1]$. Consequently, $\Pi_1$ cannot take an Arcsine distribution as in the ideal receptor case. 

\begin{figure*}
\centering
\subfigure[]{
\includegraphics[width=.482\linewidth]{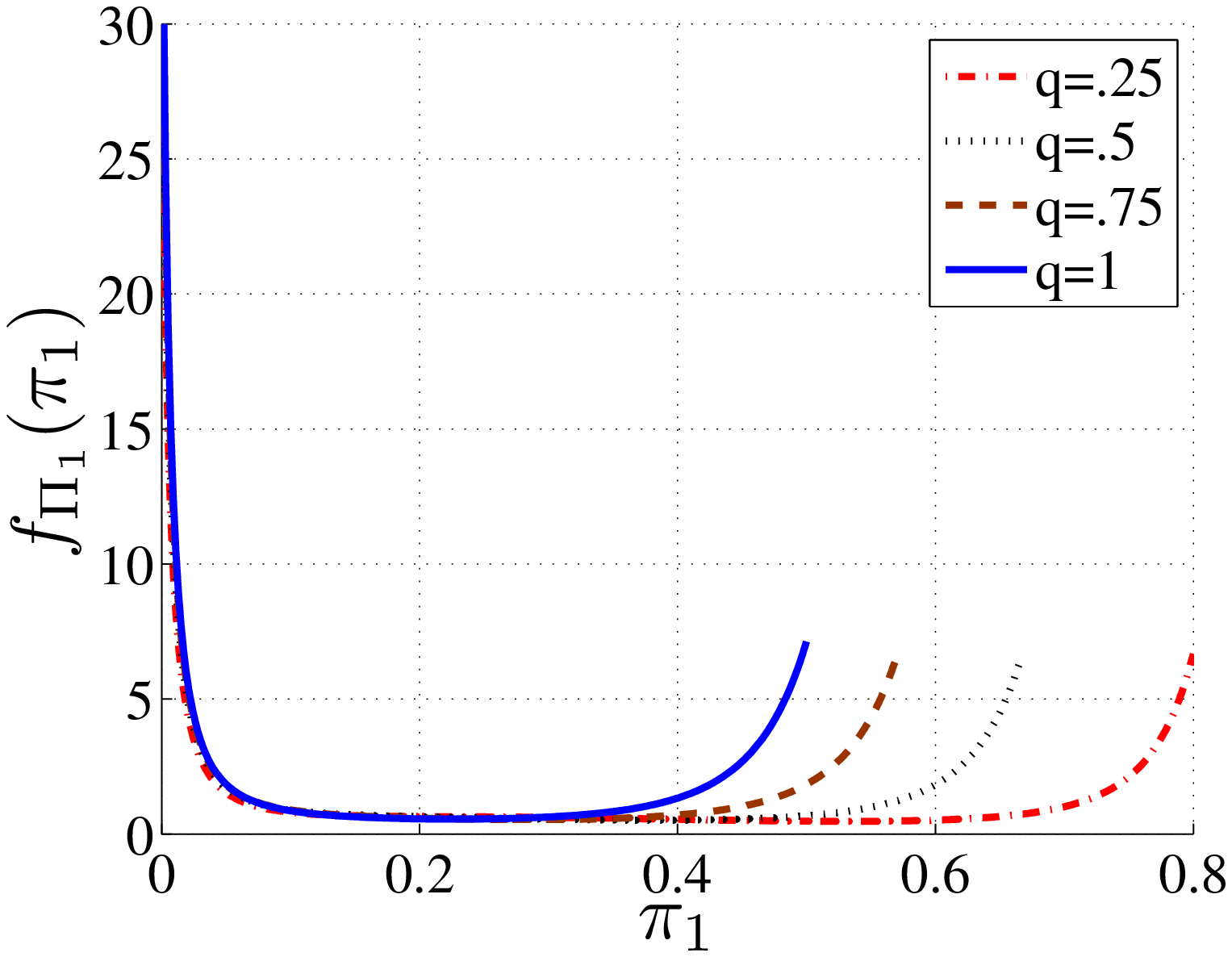}
\label{fig:markov_distribution}
}
\subfigure[]{
\includegraphics[width=.482\linewidth]{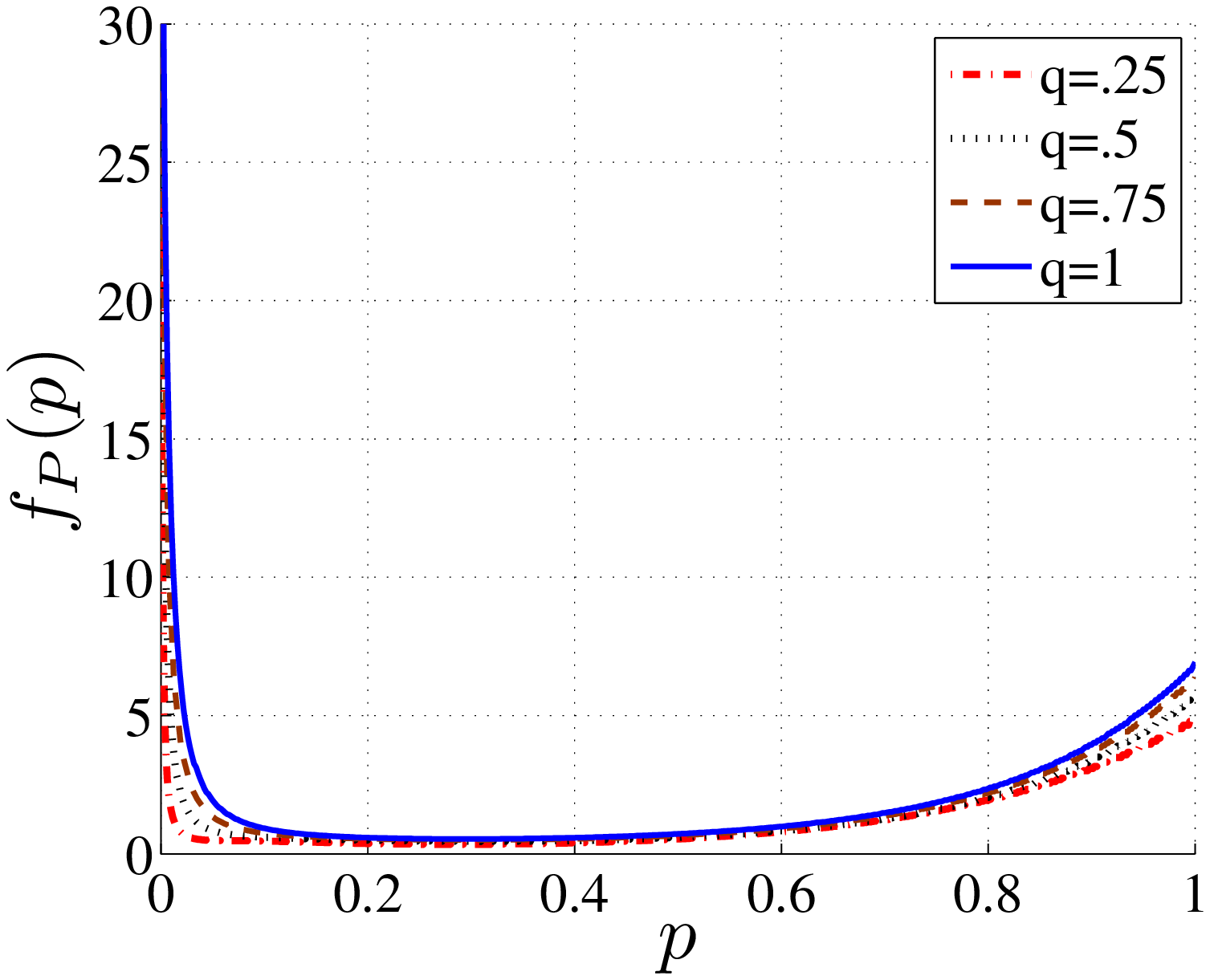}
\label{fig:markov_density}
}
\caption{ The capacity achieving distribution (a) $\pi_1$ and (b) $p$ for different values of $q$.}
\end{figure*}

One approach for finding the distribution of $\Pi_1$ is to use Lagrange multiplier to maximize~\eqref{eq:mutual2} with the constraint $\int_0^{1/(1+q)} f_{\Pi_1}(\pi_1)=1$.
Since the Lagrange method is intractable, instead, the capacity achieving distribution for $\Pi_1$ is numerically calculated by Blahut-Arimoto algorithm~\cite{Cover1990} for several values of $q$. The result is shown in Fig.~\ref{fig:markov_distribution}. We observe that the distribution of $\Pi_1$ is not symmetric and its range would decrease by increasing the parameter $q$.
From the distribution of $\Pi_1$, we can determine the optimal distribution for $P$; which in turn will determine the required distribution of the transmitter output. The distribution for $P$ is shown in Fig.~\ref{fig:markov_density}. We observe that the distribution of $P$ does not depend considerably on the value of $q$. Since the transformation from $p$ to $\pi_1$ is one-to-one and deterministic, we can calculate the capacity of the ligand receptor receiver by computing the mutual information between $\Pi_1$ and $\hat{\Pi}_1$. The numerical results for different number of receptors and $q$ are shown in Fig.~\ref{fig:capacity_plot}. We observe that the capacity is increased when $q$ in decreased. It can be explained by the fact that, except for the case of $q=0$ which MC would be meaningless, the range of $\pi_1$ increases when $q$ decreases and hence, the mutual information is increased.

\label{sec:overall}

\section{Conclusion and Discussion}
\label{sec:conclusion}
This paper is motivated by the fundamental limits on the transmission rate of information in the diffusion-based molecular channel with ligand receptors. The problem may be divided into two components. The first one corresponds to the fundamental limit on the information transmission rate due to the inherent memory in the diffusion channel which was discussed in~\cite{ISIT2011_Arash}. The second bottle-neck is at the receiver due to limits on concentration estimation by the receiver.

In our previous work, we had studied a discrete diffusion channel model in which the transmitter maps the binary bits to low and high concentration levels. Furthermore, due to the memory of the channel, the transmitter was assumed to change the duration for transmitting each bit depending on the immediate sent bit. This dependency results in four symbols with different durations and the capacity of this channel was calculated by using Shannon's model for discrete noiseless system. In this paper, we presented two models accounting for the uncertainty in estimation of concentration of molecules by the ligand receptors. Using these models, we obtained the rate limit due to the ligand receptors. The capacity achieving distribution for input was obtained for our model which showed that in order to achieve higher rates, the transmitter should mostly produce concentrations which are either low or high. It means that medium levels of concentrations should be avoided which supports our assumption for binary transmission. 

Therefore, the results from the analysis of the receiver are consistent with the transmitter model we used in~\cite{ISIT2011_Arash} to compute the capacity of the diffusion channel. The combination of results for these two components, i.e., the diffusion channel and the ligand receptor, gives a full model for diffusion-based molecular communication and cascading the two components gives an upper bound for overall information transmission rate of it. 

\balance
\bibliographystyle{IEEEtran}
\bibliography{ITW2011}

\end{document}